\documentclass[11pt]{article}
\usepackage{amsfonts}
\usepackage{}
\usepackage{bbm}

\usepackage{amsfonts, amssymb}
\usepackage{amsmath}
\usepackage{verbatim}
\usepackage{indentfirst}
\usepackage{color}

\begin{document}
\date{}
\title{{\Large\bf  3-Lie Bialgebras
\thanks{\small {Project partially supported by NSF(10871192) of
China, NSF(A2010000194) of Hebei Province, China.  Email address:
bairp1@yahoo.com.cn, lijiaqianjiayou@163.com,
mengwei139@yahoo.com.cn} }}}
\author{\small Ruipu  Bai,   Jiaqian Li, Wei Meng
\\ \small ( College of Mathematics and Computer Science,
Hebei University, Baoding 071002, China ) }

\maketitle

\vspace{2mm}\noindent{\small {\bf Abstract } $3$-Lie algebras  have
close relationships with many important fields in mathematics and
mathematical physics. The paper concerns $3$-Lie algebras. The
concepts of $3$-Lie coalgebras and $3$-Lie bialgebras are given. The
structures of such categories of  algebras, and the relationships
with $3$-Lie algebras are studied. And the classification of
$3$-dimensional $3$-Lie coalgebras and $3$-Lie bialgebras over an
algebraically closed field of characteristic zero are provided.

\vspace{1mm}\noindent{\bf Key words:}  $3$-Lie algebra; $3$-Lie
coalgebra; $3$-Lie bialgebra

 \vspace{1mm}\noindent{\bf 2000 MR subject classification}:
17B05 ~17D99}

\vspace{4mm}\noindent{\bf 1. Introduction }

The notion of $n$-Lie algebra  was introduced by Filippov in 1985
(cf \cite {F}). $n$-Lie algebras are a kind of multiple algebraic
systems appearing in many fields in mathematics and mathematical
physics. Specially, the structure of $3$-Lie algebras is applied to
the study of the supersymmetry and gauge symmetry transformations of
the world-volume theory of multiple coincident M2-branes; the Eq.
(2.2) for a $3$-Lie algebra is essential to define the action with
$N=8$ supersymmetry, and it can be regarded as a generalized Plucker
relation in the physics literature, and so on (cf.
\cite{BL,HHM,HCK,G,P}). Since the $n$-ary ($n\geq 3$)
multiplication, the structure of $n$-Lie algebras (cf
\cite{K,L,Pa,Pb,YLZ,BSZ,BBW,BSZL}, etc.) is more complicated than
that of Lie algebras. We need to excavate more constructions of
$n$-Lie algebras, and more relationships with algebraic systems
related to $n$-Lie algebras, such as Lie algebras. Note that
concepts of Lie coalgebras  and Lie bialgebras (cf \cite{M,BD,VD})
are important concepts in Lie algebras. For example, the coboundary
Lie bialgebra associates to a solution of the Classical Yang-Baxter
Equation, and it has been playing an important role in mathematics
and physics. Motivated by this, we define the $3$-Lie coalgebra and
the $3$-Lie bialgebra, and study the structure of them and the
relationships  with $3$-Lie algebras. We also classify the
$3$-dimensional $3$-coalgebras and $3$-Lie bialgebras.

Throughout this paper, all algebras are over a field $F$ of
characteristic zero. Any bracket which is not listed in the
multiplication table of a $3$-Lie algebra  or a $3$-Lie coalgebra is
assumed to be zero.

\vspace{4mm}\noindent{\bf 2. $3$-Lie coalgebras   }

 A $3$-Lie algebra $(L,\mu)$ is a vector space $L$
endowed with a $3$-ary linear skewsymmetric multiplication $\mu$
satisfying the Jacobi identity: for every $  x, y, z, u, v\in L,$

\vspace{2mm}\noindent$
  \mu(u, v, \mu(x, y, z))=\mu(x, y, \mu(u, v, z))+\mu(y, z, \mu(u, v, x))+\mu(z, x, \mu( u, v, y)).
  $

 \vspace{2mm} We also describe the $3$-Lie algebras as follows. A $3$-Lie algebra $(L,\mu)$ is a vector space $L$
endowed with a $3$-ary  linear multiplication $\mu: L\otimes
L\otimes L\rightarrow L$ satisfying

 \vspace{2mm}\noindent $\mu(1-\tau)=0$,\hfill(2.1)

\vspace{2mm}\noindent  $\mu(1\otimes1\otimes \mu)(1-\omega_1-\omega_2-\omega_3)=0, $\hfill(2.2)

 \vspace{2mm}\noindent where  $\tau: L\otimes L\otimes L\rightarrow L\otimes L\otimes
 L$, for $ ~ \forall x_1, x_2, x_3\in L, $

 \vspace{2mm}\noindent$\tau(x_1\otimes x_2\otimes
x_3)=sign(\sigma)x_{\sigma(1)}\otimes x_{\sigma(2)}\otimes
x_{\sigma(3)},\sigma\in S_3,$

 \vspace{2mm}\noindent$ 1, ~
\omega_i: L\otimes L\otimes L\otimes L\otimes L\rightarrow L\otimes
L\otimes L\otimes L\otimes L, ~ 1\leq i\leq 3$,

 \vspace{2mm}\noindent $1(x_1\otimes x_2\otimes x_3\otimes x_4\otimes x_5)=x_1\otimes x_2\otimes x_3\otimes x_4\otimes x_5,$

 \vspace{2mm}\noindent $\omega_1(x_1\otimes x_2\otimes x_3\otimes x_4\otimes x_5)=x_3\otimes x_4\otimes x_1\otimes x_2\otimes x_5,$ \hfill(2.3)

 \vspace{2mm}\noindent $\omega_2(x_1\otimes x_2\otimes x_3\otimes x_4\otimes x_5)=x_4\otimes x_5\otimes x_1\otimes x_2\otimes x_3,$ \hfill(2.4)

 \vspace{2mm}\noindent $\omega_3(x_1\otimes x_2\otimes x_3\otimes x_4\otimes x_5)=x_5\otimes x_3\otimes x_1\otimes x_2\otimes x_4.$ \hfill(2.5)

Let $(L, \mu)$ be a $3$-Lie algebra, $L^*$ be the dual space of $L$,
then we get the dual mapping $\mu^*: L^*\rightarrow L^*\otimes
L^*\otimes L^*$ of $\mu$, satisfying for every $x, y, z\in L$ and
$\xi, \eta, \zeta\in L^*, $

\vspace{2mm}\noindent  $\langle\mu^*(\xi), x\otimes y\otimes
z\rangle=\langle\xi,\mu( x, y, z)\rangle,$ \hfill(2.6)

\vspace{2mm}\noindent $ \langle\xi\otimes \eta\otimes \zeta,
x\otimes y\otimes z\rangle=\langle\xi, x\rangle\langle\eta,
y\rangle\langle\zeta, z\rangle,$\hfill(2.7)

\vspace{2mm}\noindent where $\langle, \rangle$ is the  natural
nondegenerate symmetric bilinear form on the vector space $L\oplus
L^*$ is defined by ~ $ \langle\xi, x\rangle=\xi(x),~ \xi\in L^*,
x\in L.$

Then by the definition of $3$-Lie algebras, we have ~
$Im(\mu^*)\subseteq L^*\wedge L^*\wedge L^*,$ and

\vspace{2mm}\noindent $(1-\omega_1-\omega_2-\omega_3)(1\otimes
1\otimes \mu^*)\mu^*=0,$ that is,
 for every $x, y, z, u, v\in L$ and $\xi, \eta, \zeta, \alpha, \beta\in L^*, $

\vspace{2mm}\noindent $\langle (1-\omega_1-\omega_2-\omega_3)(1\otimes 1\otimes \mu^*)\mu^*(\xi), ~ x\otimes y\otimes z\otimes  u\otimes  v \rangle$

\vspace{2mm}\noindent $=\langle (1\otimes 1\otimes \mu^*)\mu^*(\xi), ~ (1-\omega_1-\omega_2-\omega_3)(x\otimes y\otimes z\otimes  u\otimes  v) \rangle$

\vspace{2mm}\noindent $=\langle \mu^*(\xi),~  (1\otimes 1\otimes \mu)(1-\omega_1-\omega_2-\omega_3)(x\otimes y\otimes z\otimes  u\otimes  v) \rangle$

\vspace{2mm}\noindent $=\langle \xi, ~\mu(1\otimes 1\otimes
\mu)(1-\omega_1-\omega_2-\omega_3)(x\otimes y\otimes z\otimes
u\otimes  v) \rangle.$\hfill(2.8)

By the above discussion, we give the definition of $3$-Lie coalgebra.

\vspace{2mm}\noindent {\bf Definition 2.1} A 3-Lie coalgebra
$(L,\Delta)$ is a vector space $L$ with a linear mapping $\Delta:
L\rightarrow L\otimes L\otimes L$ satisfying

\vspace{2mm}\noindent(1) $Im(\Delta)\subset L\wedge L\wedge L,$
\hfill(2.9)

\vspace{2mm}\noindent (2)
$(1-\omega_1-\omega_2-\omega_3)(1\otimes1\otimes\Delta)\Delta =0,$
\hfill(2.10)

\vspace{2mm}\noindent  where $\omega_1, \omega_2, \omega_3:
L^{\otimes 5}\rightarrow L^{\otimes 5}$ satisfying identities (2.3),
(2.4) and (2.5), respectively, and $1$ is the identity mapping of
$L^{\otimes 5}$.

\vspace{2mm} Now we study  $3$-Lie coalgebras by means of structural
constants. Let $(L, \Delta)$ be a $3$-Lie coalgebra with a basis
$e_1, \cdots, e_m$. Assume

\vspace{2mm}\noindent $ \Delta(e_{l})=\sum\limits_{1\leq r<s<t\leq
n}c^{l}_{rst}e_{r}\wedge e_{s}\wedge e_{t},~ c^{k}_{rst}\in F, ~
1\leq l\leq m.$  ~~ \hfill(2.11)

\vspace{2mm}\noindent Then we have

\vspace{2mm}\noindent  $(1\otimes1\otimes\Delta)\circ\Delta
e_l=\sum\limits_{r<s<t}c_{rst}^le_{r}\wedge e_{s}\wedge \Delta
e_{t}=\sum\limits_{r<s<t }\sum\limits_{i<j<k}c_{ijk}^t c^{l}_{rst}
e_{r}\wedge e_{s}\wedge e_{i}\wedge e_{j}\wedge e_{k},$\hfill(2.12)

\vspace{3mm}\noindent
$(1-\omega_1-\omega_2-\omega_3)\circ(1\otimes1\otimes\Delta)\circ\Delta
e_l$

\vspace{3mm}\noindent
$=(1-\omega_1-\omega_2-\omega_3)(\sum\limits_{r<s<t
}\sum\limits_{i<j<k}c_{ijk}^t c^{l}_{rst}e_{r}\wedge e_{s}\wedge
e_{i}\wedge e_{j}\wedge e_{k})$

\vspace{3mm}\noindent $=\sum\limits_{r<s<t }\sum\limits_{i<j<k}
c_{ijk}^t c^{l}_{rst}e_{r}\wedge e_{s}\wedge e_{i}\wedge e_{j}\wedge
e_{k}-\sum\limits_{r<s<t }\sum\limits_{i<j<k}c_{rsk}^t c^{l}_{ijt}
e_{i}\wedge e_{j}\wedge e_{r}\wedge e_{s}\wedge e_{k}$

\vspace{3mm}\noindent $-\sum\limits_{r<s<t
}\sum\limits_{i<j<k}c_{jkt}^l c^{t}_{rsi}e_{j}\wedge e_{k}\wedge
e_{r}\wedge e_{s}\wedge e_{i}-\sum\limits_{r<s<t
}\sum\limits_{i<j<k}c_{kit}^l c^{t}_{rsj} e_{k}\wedge e_{i}\wedge
e_{r}\wedge e_{s}\wedge e_{j}$

\vspace{3mm}\noindent $=\sum\limits_{r<s<t
}\sum\limits_{i<j<k}[c_{ijk}^t c^{l}_{rst}-c_{rsk}^t
c^{l}_{ijt}-c_{jkt}^l c^{t}_{rsi}-c_{kit}^l c^{t}_{rsj}]e_{r}\wedge
e_{s}\wedge e_{i}\wedge e_{j}\wedge e_{k},$  \hfill(2.13)

\vspace{3mm}\noindent
$c^{k}_{i_1i_2i_3}=sgn(\sigma)c^{k}_{i_{\sigma(1)}i_{\sigma(2)}i_{\sigma(3)}},
~ 1\leq i_1, i_2, i_3\leq m.$  \hfill(2.14)

\vspace{2mm}\noindent Therefore,

\vspace{2mm}\noindent$\sum\limits_{k=1}^{n}(c_{ijk}^t
c^{l}_{rst}-c_{rsk}^t c^{l}_{ijt}-c_{jkt}^l c^{t}_{rsi}-c_{kit}^l
c^{t}_{rsj})=0, ~ 1\leq i, j, k, l\leq m.$\hfill(2.15)

\vspace{2mm} Following the above discussions, we obtain the
structural description of $3$-Lie coalgebras in terms of structural
constants.

\vspace{2mm}\noindent{\bf Theorem 2.2 } Let $L$ be an
$n$-dimensional vector space with a basis $e_1, \cdots, e_m$,
$\Delta: L\rightarrow L\otimes L\otimes L$ be defined as (2.11).
Then $(L, \Delta)$ is a $3$-Lie coalgebra if and only if the
constants $c^l_{ijk}$, $1\leq i, j, k\leq m$ satisfy identities
(2.14) and (2.15). $\Box$

\vspace{2mm} Now let  $(L, \mu)$ be  a 3-Lie algebra with a basis
$e_1, e_2, \cdots, e_m$, and the the multiplication of $L$ in the
basis is as follows

\vspace{2mm}\noindent $\mu(e_{i}, e_{j},
e_{k})=\sum\limits_{l=1}^{n}c^{l}_{ijk}e_l, ~
 c^{l}_{ijk}\in F, ~1\leq i, j, k, l \leq m.$ \hfill(2.16)

\vspace{2mm}\noindent By the skew-symmetry of the multiplication of
 $3$-Lie algebras,  we have

\vspace{2mm}\noindent$c^{l}_{i_1i_2i_3}$
$=sgn(\sigma)c^{l}_{i_{\sigma(1)}i_{\sigma(2)}i_{\sigma(3)}},$

\vspace{2mm}\noindent that is, $c^{l}_{i_1i_2i_3}$, $1\leq l, i_1,
i_2, i_3\leq m$ satisfy the identity (2.14). By the Jacobi identity
(2.2)

\vspace{2mm}\noindent$\mu((e_i,$ $e_j,$ $e_k), e_s,$ $e_t)=$
$\mu(\mu( e_i,e_s,e_t), e_j,e_k)$ $+\mu( e_i, $ $\mu(e_j,e_s,e_t),
e_k)$ $+\mu(e_i, e_j,$ $\mu(e_k,e_s,e_t)).$

\vspace{2mm}\noindent  Since

\vspace{2mm}\noindent $\mu(\mu
(e_i,e_j,e_k),e_s,e_t)=\mu(\sum\limits_{l=1}^{n}c^{l}_{ijk}e_l,e_s,e_t)=\sum\limits_{l=1}^{n}c^{l}_{ijk}\mu(e_l,e_s,e_t)=\sum\limits_{r=1}^{n}\sum\limits_{l=1}^{n}c^{l}_{ijk}c^{r}_{lst}e_r,$

\vspace{2mm}\noindent
$\mu(\mu(e_i,e_s,e_t),e_j,e_k)+\mu(e_i,(e_j,e_s,e_t),e_k)+\mu(e_i,e_j,\mu(e_k,e_s,e_t))$

\vspace{2mm}\noindent$=\mu(\sum\limits_{l=1}^{n}c^{l}_{ist}e_l,e_j,e_k)+\mu(e_i,\sum\limits_{l=1}^{n}c^{l}_{jst}e_l,e_k)$
$+\mu(e_i,e_j,\sum\limits_{l=1}^{n}c^{l}_{kst}e_l)$

\vspace{2mm}\noindent$=\sum\limits_{r=1}^{n}\sum\limits_{l=1}^{n}c^{l}_{ist}c^{r}_{ljk}e_r+\sum\limits_{r=1}^{n}\sum\limits_{l=1}^{n}c^{l}_{jst}c^{r}_{ilk}e_r+
\sum\limits_{r=1}^{n}\sum\limits_{l=1}^{n}c^{l}_{kst}c^{r}_{ijl}e_r,$

\vspace{2mm}\noindent we have
$\sum\limits_{l=1}^{n}(c^{l}_{ijk}c^{r}_{lst}+c^{r}_{lkj}c^{l}_{ist}+
c^{r}_{lik}c^{l}_{jst}+c^{r}_{lji}c^{l}_{kst})=0,$ that is,
$\{c^{l}_{i_1i_2i_3}, 1\leq i_1, i_2, i_3\leq n\}$ satisfies the
identity (2.15).

Let $L^{\ast}$ be  the dual space of $L$, $e^{1}, \cdots, e^{m}$ be
the dual basis of $e_1, \cdots, e_m$, that is, $\langle e^{i},
e_j\rangle=\delta_{ij}, 1\leq i, j\leq m.$ Assume
$\mu^*:L^*\rightarrow L^*\otimes L^*\otimes L^*$ is the dual mapping
of $\mu$ defined by (2.6), that is for every $\xi\in L^*, x, y, z\in
L$,~$\langle \mu^*(\xi), x\otimes y\otimes z \rangle=\langle\xi,
\mu(x, y, z)\rangle.$

\vspace{2mm}\noindent Then for every $~ 1\leq l\leq m,$ we have

\vspace{2mm}\noindent $\mu^*(e^{l})=\sum\limits_{1\leq i<j<k\leq
n}c^{l}_{ijk}e^{i}\wedge e^{j}\wedge e^{k}.$ \hfill(2.17)

\vspace{2mm}\noindent Follows the identities (2.11) and (2.15),
$(L^*, \mu^*)$ is a $3$-Lie coalgebra.

Conversely, if $(L, \Delta)$ is a $3$-Lie coalgebra with a basis
$e_1, \cdots, e_m$ satisfying (2.11), $L^*$ is the dual space of $L$
with the dual basis $e^1, \cdots, e^m$. Then the dual mapping
$\Delta^*: L^*\otimes L^*\otimes L^*\rightarrow L^*$ of $\Delta$
satisfies, for every $\xi, \eta, \zeta\in L^*, x\in L,$

\vspace{2mm}\noindent $\langle \Delta^*(\xi, \eta, \zeta),
x\rangle=\langle \xi\otimes \eta \otimes \zeta,
\Delta(x)\rangle.$\hfill(2.18)

\vspace{2mm}\noindent Then  $\Delta^*(e^{i}, e^{j},
e^{k})=\sum\limits_{l=1}^{n}c^{l}_{ijk}e^l, ~
 c^{l}_{ijk}\in F, ~1\leq i, j, k, l \leq m$ and $\Delta^*$ satisfies  identity (2.15).

Summarizing above discussions, we have the following result.

\vspace{2mm}\noindent{\bf Theorem 2.3 } Let $L$ be a vector space,
$\Delta: L\rightarrow L\otimes L\otimes L$. Then  $(L, \Delta)$ is a
$3$-Lie coalgebra if and only if $(L^*, \Delta^*)$ is a $3$-Lie
algebra. ~~$\Box$

We can also give an equivalence description of Theorem 2.3.

\vspace{2mm}\noindent{\bf Theorem 2.4} Let $L$ be a vector space
over a field $F$, and   $\mu: L\otimes L\otimes L\rightarrow L$ be a
$3$-ary linear mapping. Then $(L, \mu)$ is a $3$-Lie algebra with
the multiplication $\mu$ if and only if $(L^*, \mu^*)$ is a $3$-Lie
coalgebra with  $\mu^*: L^*\rightarrow L^*\otimes L^*\otimes L^*$,
where $\mu^*$ is the dual mapping of $\mu$.  $\Box$

\vspace{2mm}\noindent{\bf Example 2.5} Let $L$ be a $5$-dimensional
$3$-Lie algebra with the following multiplication

\vspace{2mm}\noindent$\mu(e_2, e_3, e_4)=e_1$, $\mu(e_3, e_4,
e_5)=e_3+2e_2$, $\mu(e_2, e_4, e_5)=e_3$, $\mu(e_1, e_4, e_5)=e_1$,

\vspace{2mm}\noindent where $e_1, e_2, e_3, e_4, e_5$ is a basis of
$L$. By Theorem 2.3, $(L^*, \mu^*)$ is a $3$-Lie coalgebra with the
linear mapping $\mu^*: L^*\rightarrow L^*\otimes L^*\otimes L^*$
satisfying

\vspace{2mm}\noindent $\mu^*(e^1)=e^2\wedge e^3\wedge e^4 +
e^1\wedge e^4\wedge e^5$, $\mu^*(e^3)=e^3\wedge e^4\wedge
e^5+e^2\wedge e^4\wedge e^5$, $\mu^*(e^2)=\alpha e^3\wedge e^4\wedge
e^5, $

\vspace{2mm}\noindent where $e^1, e^2, e^3, e^4, e^5\in L^*$ is  the
dual basis of  $e_1, e_2, e_3, e_4, e_5$. $\Box$

\vspace{2mm}\noindent{\bf Example 2.6} Let $(L, \Delta)$ be a
$4$-dimensional $3$-Lie coalgebra with a basis $e_1, e_2, e_3, e_4$,
and satisfying $\Delta (e_2)=\alpha e_3\wedge e_1 \wedge e_4, \Delta
(e_3)=e_3\wedge e_1 \wedge e_4 + e_2\wedge e_1 \wedge e_4.$

 $(L^*, \Delta^*)$ is a $3$-Lie algebra with
the $3$-ary linear skew-symmetric mapping $\Delta^*: L^*\otimes
L^*\otimes L^*\rightarrow L^*$ satisfying

\vspace{2mm}\noindent $\Delta^*(e^2, e^3, e^4)=\alpha e^2 + e^3,$
$\Delta^*(e^2, e^1, e^4)=e^3$, where $e^1, e^2, e^3, e^4\in L^*$ is
the dual basis of  $e_1, e_2, e_3, e_4$. $\Box$

\vspace{2mm}\noindent{\bf Definition 2.7 } Let $(L_1, \Delta_1)$ and
$(L_2, \Delta_2)$ be $3$-Lie coalgebras. If there is a linear
isomorphism $\varphi: L_1\rightarrow L_2$  satisfying

\vspace{2mm}\noindent$(\varphi\otimes \varphi\otimes
\varphi)(\Delta_1(e))=\Delta_2(\varphi(e))$, for every $e\in L_1$,
\hfill(2.19)

\vspace{2mm}\noindent then $(L_1, \Delta_1)$ is  isomorphic to
$(L_2, \Delta_2)$, and $\varphi$ is called a $3$-Lie coalgebra
 isomorphism,  where

\vspace{2mm}\noindent $(\varphi\otimes \varphi\otimes
\varphi)\sum\limits_i(a_i\otimes  b_i\otimes
c_i)=\sum\limits_i\varphi(a_i)\otimes \varphi(b_i)\otimes
\varphi(c_i)$.\hfill(2.20)

\vspace{2mm}\noindent{\bf Theorem 2.8 } Let $(L_1, \Delta_1)$ and
$(L_2, \Delta_2)$ be $3$-Lie coalgebras. Then $\varphi:
L_1\rightarrow L_2$ is a $3$-Lie coalgebra isomorphism from $(L_1,
\Delta_1)$  to $(L_2, \Delta_2)$ if and only if the dual mapping
$\varphi^*: L_2^*\rightarrow L_1^*$ is a $3$-Lie algebra isomorphism
from $(L_2^*, \Delta_2^*)$ to $(L_1^*, \Delta^*_1)$, where for every
$\xi\in L_2^*, v\in L_1$, $\langle \varphi^*(\xi),
v\rangle=\langle\xi, \varphi(v)\rangle$.

\vspace{2mm}\noindent{\bf Proof } Since $(L_1, \Delta_1)$ and $(L_2,
\Delta_2)$ are $3$-Lie coalgebras, $(L^*_1, \Delta^*_1)$ and
$(L^*_2, \Delta^*_2)$ are $3$-Lie algebras.  Let $\varphi:
L_1\rightarrow L_2$ be a $3$-Lie coalgebra isomorphism from $(L_1,
\Delta_1)$  to $(L_2, \Delta_2)$. Then the dual mapping $\varphi^*:
L_2^*\rightarrow L_1^*$ is a linear isomorphism. And for every $\xi,
\eta, \zeta\in L^*_2, x\in L_1^*,$ by identities (2.6) and  (2.18)

\vspace{2mm}\noindent $\langle \varphi^*\Delta_2^*(\xi, \eta,
\zeta), x\rangle=\langle\Delta_2^*(\xi, \eta, \zeta), \varphi(x)
\rangle$

\vspace{2mm}\noindent $=\langle \xi\otimes\eta\otimes \zeta),
\Delta_2(\varphi(x))\rangle=\langle \xi\otimes \eta\otimes \zeta,
(\varphi\otimes \varphi\otimes \varphi)\Delta_1 (x) \rangle$

\vspace{2mm}\noindent $=\langle \varphi^*(\xi)\otimes
\varphi^*(\eta)\otimes \varphi^*(\zeta), \Delta_1 (x) \rangle$

\vspace{2mm}\noindent $=\langle \Delta_1^*(\varphi^*(\xi),
\varphi^*(\eta), \varphi^*(\zeta)), x \rangle.$

\vspace{2mm}\noindent It follows $\varphi^*\Delta_2^*(\xi, \eta,
\zeta)=\Delta_1^*(\varphi^*(\xi), \varphi^*(\eta),
\varphi^*(\zeta))$, that is, $\varphi^*: L_2^*\rightarrow L_1^*$ is
a $3$-Lie algebra isomorphism from $(L_2^*, \Delta_2^*)$ to $(L_1^*,
\Delta^*_1)$.

Similarly, the conversion  holds. ~~$\Box$

\vspace{2mm} \noindent {\bf  3.  $3$-Lie bialgebras}

 In this section we define $3$-Lie
bialgebra, which  is an algebraic system with $3$-ary multiplication
structures of $3$-Lie algebra and  $3$-Lie coalgebra, simultaneouly.
We first give the definition.

\vspace{2mm}\noindent {\bf Definition 3.1} A 3-Lie bialgebra  is a
triple $(L, \mu, \Delta)$ such that

\vspace{2mm}\noindent (1) $(L, \mu)$ is a $3$-Lie algebra with  the
multiplication $\mu: L\wedge L\wedge L\rightarrow L,$

\vspace{2mm}\noindent (2) $(L, \Delta)$ is a $3$-Lie coalgebra with
$\Delta: L\rightarrow L\wedge L\wedge L$,

\vspace{2mm}\noindent (3) $\Delta$ and $\mu$ satisfy the following
condition

\vspace{2mm}\noindent$\Delta\mu(x, y,
z)=ad_{\mu}^{(3)}(x,y)\Delta(z)+ad_{\mu}^{(3)}(y,z)\Delta(x)+ad_{\mu}^{(3)}(z,x)\Delta(y),$\hfill(3.1)

\vspace{2mm}\noindent  where  $ad_{\mu}(x, y): L\wedge L\rightarrow
End(L), ad_{\mu}(x, y)(z)=\mu(x, y, z)$ for $x, y, z\in L$;

\vspace{2mm}\noindent $ad_{\mu}^{(3)}(x, y),$ $ ad_{\mu}^{(3)}(z,
x),$ $ ad_{\mu}^{(3)}(y, z):$ $L\otimes L\otimes L\rightarrow
L\otimes L\otimes L$ are $3$-ary linear mappings satisfying for
every $u, v, w\in L$

 \vspace{2mm}\noindent $ad_{\mu}^{(3)}(x, y)(u\otimes v\otimes w)=(ad_{\mu}(x,
y)\otimes 1\otimes 1)(u\otimes v\otimes w)$

 \vspace{2mm}\noindent$+(1\otimes ad_{\mu}(x,
y)\otimes 1)(u\otimes v\otimes w)+(1\otimes 1\otimes ad_{\mu}(x,
y))(u\otimes v\otimes w)$

\vspace{2mm}\noindent$ =\mu(x, y, u)\otimes v\otimes w+ u\otimes
\mu(x, y, v)\otimes w+u\otimes v\otimes \mu(x, y, w),$\hfill(3.2)

\vspace{2mm}\noindent  and the similar action for $ad_{\mu}^{(3)}(z,
x)$ and $ad_{\mu}^{(3)}(y, z).$ $\Box$

The condition (1) is equivalent to that  $(L^*, \mu^*)$ is a $3$-Lie
coalgebra with $\mu^*: L^*\rightarrow L^*\wedge L^*\wedge L^*$
defined by (2.6).

The condition (2) is equivalent to that$(L^*, \Delta^*)$ is a
$3$-Lie algebra with the $3$-Lie bracket $\Delta^*: L^*\wedge
L^*\wedge L^*\rightarrow L^*$ defined by (2.18) satisfying for every
permutation $\sigma\in S_3$, and $\xi_1, \cdots, \xi_5\in L^*,$

\vspace{2mm}\noindent $\Delta^*(\xi_1, \xi_2,
\xi_3)=sign(\sigma)\Delta^*(\xi_{\sigma(1)}, \xi_{\sigma(2)},
\xi_{\sigma(3)})$,

\vspace{2mm}\noindent $\Delta^*(\xi_1, \xi_2, \Delta^*(\xi_3, \xi_4,
\xi_5))$

\vspace{2mm}\noindent$=\Delta^*(\xi_4, \xi_5, \Delta^*(\xi_1, \xi_2,
\xi_3))+ \Delta^*(\xi_5, \xi_3, \Delta^*(\xi_1, \xi_2,
\xi_4))+\Delta^*(\xi_3, \xi_4, \Delta^*(\xi_1, \xi_2, \xi_5)).$

An alternate way of writing the condition (3) is for every $x, y,
z\in L, \xi, \eta, \zeta\in L^*$,

\vspace{2mm}\noindent $\langle \Delta^*(\xi, \eta, \zeta), \mu(x, y,
z)\rangle =\langle \xi\otimes \eta\otimes \zeta, \Delta(\mu(x, y,
z))\rangle=\langle \xi\otimes \eta\otimes \zeta, ad_{\mu}^{(3)}(x,
y)\Delta(z)\rangle$

\vspace{2mm}\noindent $ + \langle \xi\otimes \eta\otimes \zeta,
ad_{\mu}^{(3)}(z, x)\Delta(y)\rangle+ \langle \xi\otimes \eta\otimes
\zeta,ad_{\mu}^{(3)}(y,z)\Delta(x)\rangle.$

\vspace{2mm}\noindent{\bf Example 3.2 } Let $L$ be a $4$-dimensional
vector space with a basis $e_1, e_2, e_3, e_4.$ Defines

\vspace{2mm}\noindent $\mu: L\wedge L \wedge L\rightarrow L, ~~
\begin{cases} \mu(e_1, e_3, e_4)=e_1,\\ \mu(e_2, e_3, e_4)=e_2;
\end{cases}$

\vspace{2mm}\noindent$\Delta: L\rightarrow L\wedge L \wedge L, ~~
\begin{cases}\Delta(e_1)=e_3\wedge e_2\wedge e_4,\\ \Delta(e_3)=e_1\wedge e_2\wedge
e_4.\end{cases}$~~

\vspace{2mm}\noindent Then by the direct computation,  the triple
$(L, \mu, \Delta)$ is a $3$-Lie bialgebra.  ~~$\Box$

 Let $(L, \mu)$ be a $3$-Lie
algebra, for $ \forall x, y, z\in L$,  $ad_{\mu}: L\wedge
L\rightarrow gl(L),$ $ ad_{\mu}(x, y)(z)=\mu(x, y, z),$ be the
adjoint representation of $3$-Lie algebra $L$. Then $ad^*_{\mu}:
L\wedge L\rightarrow gl(L^*),$ defined by

\vspace{2mm}\noindent $ \langle ad*_{\mu}(x, y)(\xi), z\rangle
=-\langle \xi, ad_{\mu}(x, y)(z)\rangle=-\langle \xi, \mu(x, y,
z)\rangle,$ $ \forall x, y, z\in L, \xi\in L^*$

\vspace{2mm}\noindent is the coadjoint  representation of $L$.

\vspace{2mm}\noindent {\bf Theorem 3.3 } Let $(L, \mu, \Delta)$ be a
$3$-Lie bialgebra. Then $(L^*, \Delta^*, \mu^*)$ is a $3$-Lie
bialgebra, and it is called the dual $3$-Lie bialgebra of  $(L, \mu,
\Delta)$.

\vspace{2mm}\noindent {\bf Proof } Since $(L, \mu, \Delta)$ is a
$3$-Lie bialgebra, by  Theorem 2.3 and Theorem 2.4, $(L^*,
\Delta^*)$ be a $3$-Lie algebra in the multiplication (2.18), and
$(L^*, \mu^*)$ be a $3$-Lie coalgebra in the multiplication (2.6).

We will prove that $\mu^*: L^* \rightarrow L^*\wedge L^*\wedge L^*$
satisfies identity (3.1), that is, the following identity holds for
every $\xi, \eta, \zeta\in L^*$

\vspace{2mm}\noindent$\mu^*(\Delta^*(\xi, \eta,
\zeta))=ad_{\Delta^*}^{(3)}(\xi,
\eta)\mu^*(\zeta)+ad_{\Delta^*}^{(3)}(\eta,
\zeta)\mu^*(\xi)+ad_{\Delta^*}^{(3)}(\zeta, \xi)\mu^*(\eta).$
\hfill(3.3)

\vspace{2mm} For every $x, y, z\in L$ and $\xi, \eta, \zeta\in L^*$,
by identities (2.6) and (2.18)

\vspace{2mm}\noindent$\langle\mu^*\Delta^*(\xi, \eta, \zeta),
x\otimes y\otimes z\rangle=$

\vspace{2mm}\noindent $\langle\Delta^*(\xi, \eta, \zeta), \mu(x, y,
z)\rangle=\langle\xi\otimes \eta\otimes \zeta, \Delta(\mu(x, y,
z))\rangle=$

\vspace{2mm}\noindent$\langle \xi\otimes \eta\otimes \zeta,
ad_{\mu}^{(3)}(x, y)\Delta(z)\rangle$
 $ + \langle \xi\otimes \eta\otimes \zeta,
ad_{\mu}^{(3)}(z, x)\Delta(y)\rangle+ \langle \xi\otimes \eta\otimes
\zeta,ad_{\mu}^{(3)}(y,z)\Delta(x)\rangle.$

\vspace{2mm} Without loss of generality, suppose
$\Delta(z)=\sum\limits_ia_i\otimes b_i\otimes c_i,$ ~ where $ a_i,
b_i, c_i\in L$. Then

\vspace{2mm}\noindent$\langle \xi\otimes \eta\otimes \zeta,
ad_{\mu}^{(3)}(x, y)\Delta(z)\rangle=\langle \xi\otimes \eta\otimes
\zeta, ad_{\mu}^{(3)}(x, y)(\sum\limits_ia_i\otimes b_i\otimes
c_i)\rangle=$

\vspace{2mm}\noindent$\langle \xi\otimes \eta\otimes \zeta,
\sum\limits_i(ad_{\mu}(x, y)(a_i)\otimes b_i\otimes c_i+a_i\otimes
ad_{\mu}(x, y)(b_i)\otimes  c_i+a_i\otimes b_i\otimes ad_{\mu}(x,
y)(c_i)\rangle$

\vspace{2mm}\noindent$=-\sum\limits_i\langle ad_{\mu}^*(x,
y)(\xi)\otimes \eta\otimes \zeta+\xi\otimes ad_{\mu}^*(x,
y)(\eta)\otimes \zeta+\xi\otimes \eta\otimes ad_{\mu}^*(x,
y)(\zeta), a_i\otimes b_i\otimes c_i\rangle$

\vspace{2mm}\noindent$=-\langle ad_{\mu}^*(x, y)(\xi)\otimes
\eta\otimes \zeta+\xi\otimes ad_{\mu}^*(x, y)(\eta)\otimes
\zeta+\xi\otimes \eta\otimes ad_{\mu}^*(x,
y)(\zeta),\Delta(z)\rangle$

\vspace{2mm}\noindent$=-\langle \Delta^*(ad_{\mu}^*(x, y)(\xi),
\eta, \zeta)+\Delta^*(\xi, ad_{\mu}^*(x,
y)(\eta),\zeta)+\Delta^*(\xi, \eta, ad_{\mu}^*(x,
y)(\zeta)),z\rangle$

\vspace{2mm}\noindent$=-\langle
ad_{\Delta^*}(\eta,\zeta)(ad_{\mu}^*(x,
y)(\xi)+ad_{\Delta^*}(\zeta,\xi)ad_{\mu}^*(x,
y)(\eta)+ad_{\Delta^*}(\xi, \eta)ad_{\mu}^*(x, y)(\zeta)),
z\rangle.$\hfill(3.4)

Similarly, we have

\vspace{2mm}\noindent$\langle \xi\otimes \eta\otimes \zeta,
ad_{\mu}^{(3)}(z, x)\Delta(y)\rangle $

\vspace{2mm}\noindent$=-\langle
ad_{\Delta^*}(\eta,\zeta)(ad_{\mu}^*(z,
x)(\xi)+ad_{\Delta^*}(\zeta,\xi)ad_{\mu}^*(z,
x)(\eta)+ad_{\Delta^*}(\xi, \eta)ad_{\mu}^*(z, x)(\zeta)),
y\rangle,$\hfill(3.5)

\vspace{2mm}\noindent$\langle \xi\otimes \eta\otimes
\zeta,ad_{\mu}^{(3)}(y,z)\Delta(x)\rangle$

\vspace{2mm}\noindent$=-\langle
ad_{\Delta^*}(\eta,\zeta)ad_{\mu}^*(y,
z)(\xi)+ad_{\Delta^*}(\zeta,\xi)ad_{\mu}^*(y,
z)(\eta)+ad_{\Delta^*}(\xi, \eta)ad_{\mu}^*(y, z)(\zeta)),
x\rangle.$\hfill(3.6)

\vspace{2mm}Since

\vspace{2mm}\noindent$-\langle
ad_{\Delta^*}(\eta,\zeta)ad_{\mu}^*(x, y)(\xi), z\rangle-\langle
ad_{\Delta^*}(\eta,\zeta)ad_{\mu}^*(y, z)(\xi), x\rangle-\langle
ad_{\Delta^*}(\eta,\zeta)ad_{\mu}^*(z, x)(\xi), y\rangle$

\vspace{2mm}\noindent$=\langle ad_{\mu}^*(x, y)(\xi),
ad^*_{\Delta^*}(\eta,\zeta)z\rangle+\langle ad_{\mu}^*(y, z)(\xi),
ad^*_{\Delta^*}(\eta,\zeta)x\rangle+\langle ad_{\mu}^*(z, x)(\xi),
ad^*_{\Delta^*}(\eta,\zeta)y\rangle$

\vspace{2mm}\noindent$=-\langle \xi, \mu(x, y,
ad^*_{\Delta^*}(\eta,\zeta)z)\rangle-\langle \xi,\mu(
ad^*_{\Delta^*}(\eta,\zeta)x, y, z)\rangle-\langle\xi,\mu (x,
ad^*_{\Delta^*}(\eta,\zeta)y, z)\rangle$

\vspace{2mm}\noindent$=-\langle \mu^*(\xi),
ad_{\Delta^*}^{*(3)}(\eta, \zeta)(x\otimes  y\otimes
z)\rangle=\langle ad_{\Delta^*}^{(3)}(\eta, \zeta)\mu^*(\xi),
x\otimes  y\otimes z\rangle,$ and

\vspace{2mm}\noindent$-\langle ad_{\Delta^*}(\zeta,
\xi)ad_{\mu}^*(x, y)(\eta), z\rangle-\langle ad_{\Delta^*}(\zeta,
\xi)ad_{\mu}^*(y, z)(\eta), x\rangle-\langle ad_{\Delta^*}(\zeta,
\xi)ad_{\mu}^*(z, x)(\eta), y\rangle$

\vspace{2mm}\noindent$=-\langle \mu^*(\eta),
ad_{\Delta^*}^{*(3)}(\zeta, \xi)(x\otimes  y\otimes
z)\rangle=\langle ad_{\Delta^*}^{(3)}(\zeta, \xi)\mu^*(\eta),
x\otimes y\otimes )\rangle,$ and

\vspace{2mm}\noindent$-\langle ad_{\Delta^*}(\xi, \eta)ad_{\mu}^*(x,
y)(\zeta), z\rangle-\langle ad_{\Delta^*}(\xi, \eta)ad_{\mu}^*(y,
z)(\zeta), x\rangle-\langle ad_{\Delta^*}(\xi, \eta)ad_{\mu}^*(z,
x)(\zeta), y\rangle$

\vspace{2mm}\noindent$=-\langle \mu^*(\eta),
ad_{\Delta^*}^{*(3)}(\zeta, \xi)(x\otimes  y\otimes
z)\rangle=\langle ad_{\Delta^*}^{(3)}(\zeta, \xi)\mu^*(\eta),
x\otimes  y\otimes z\rangle,$

\vspace{2mm}\noindent  the identity (3.3) holds. It follows the
result. ~~ $\Box$

By the duality property, every $3$-Lie bialgebra has dual $3$-Lie
bialgebra whose dual is the $3$-Lie bialgebra itself. And
summarizing identities (3.2) - (3.6), we obtain

\vspace{2mm}\noindent$\langle\Delta^*(\xi, \eta, \zeta), \mu(x, y,
z\rangle$

\vspace{2mm}\noindent$=-\langle
ad_{\Delta^*}(\eta,\zeta)(ad_{\mu}^*(x,
y)(\xi)+ad_{\Delta^*}(\zeta,\xi)ad_{\mu}^*(x,
y)(\eta)+ad_{\Delta^*}(\xi, \eta)ad_{\mu}^*(x, y)(\zeta)),
z\rangle-$

\vspace{2mm}\noindent$\langle
ad_{\Delta^*}(\eta,\zeta)(ad_{\mu}^*(z,
x)(\xi)+ad_{\Delta^*}(\zeta,\xi)ad_{\mu}^*(z,
x)(\eta)+ad_{\Delta^*}(\xi, \eta)ad_{\mu}^*(z, x)(\zeta)),
y\rangle-$

\vspace{2mm}\noindent$\langle ad_{\Delta^*}(\eta,\zeta)ad_{\mu}^*(y,
z)(\xi)+ad_{\Delta^*}(\zeta,\xi)ad_{\mu}^*(y,
z)(\eta)+ad_{\Delta^*}(\xi, \eta)ad_{\mu}^*(y, z)(\zeta)),
x\rangle.$\hfill(3.7)

\vspace{2mm}\noindent {\bf Example 3.4 } From  Theorem 3.3, the dual
$3$-Lie bialgebra $(L^*, \Delta^*, \mu^*)$ of $(L, \mu, \Delta)$ in
Example 3.2  satisfying $\langle e^i, e_j\rangle=\delta_{ij}, 1\leq
i, j\leq 4$, and

\vspace{2mm}\noindent $\mu^*: L^*\rightarrow L^*\wedge L^* \wedge
L^*, ~~
\begin{cases} \mu^*(e^1)=e^1\wedge e^3\wedge e^4,\\ \mu^*(e^2)=e^2\wedge e^3\wedge e^4;
\end{cases}$

\vspace{2mm}\noindent$\Delta^*:  L^*\wedge L^* \wedge L^*\rightarrow
L^*, ~~
\begin{cases}\Delta^*(e^3\wedge e^2\wedge e^4)=e^1,\\ \Delta^*(e^1\wedge e^3\wedge
e^4)=e^3.\end{cases}$~~ $\Box$

\vspace{2mm} At last of this section we describe $3$-Lie bialgebras
by the structural constants.

Let $(L, \mu, \Delta)$ be a $3$-Lie bialgebra with the
multiplications in the basis $e_1, \cdots, e_m$ as follows

\vspace{2mm}\noindent$\mu(e_i, e_j,
e_k)=\sum\limits_{l=1}^mc_{ijk}^le_l, ~~
\Delta(e_l)=\sum\limits_{1\leq i<j<k\leq m}a^{ijk}_le_i\wedge
e_j\wedge e_k, $\hfill(3.8)

\vspace{2mm}\noindent where $c_{ijk}^l, a^{ijk}_l\in F$ $~ 1\leq i<
j<k\leq m,  1\leq l\leq m, $ satisfy identities (2.14) and (2.15),
respectively. Then

\vspace{2mm}\noindent$\Delta\mu(e_i, e_j, e_k)=\sum\limits_{l=1}^m
c_{ijk}^l\Delta(e_l)=\sum\limits_{l=1}^m\sum\limits_{
r<s<t}c_{ijk}^la^{rst}_le_r\wedge e_s\wedge e_t, 1\leq i< j<k\leq
m.$\hfill(3.9)

\vspace{2mm}\noindent By identity (3.1),

\vspace{2mm}\noindent$\Delta\mu(e_i, e_j, e_k)=ad_{\mu}^{(3)}(e_i,
e_j)\Delta(e_k)+ad_{\mu}^{(3)}(e_j,
e_k)\Delta(e_i)+ad_{\mu}^{(3)}(e_k,
e_i)\Delta(e_j)$

\vspace{2mm}\noindent$=\sum\limits_{r<s<t}a^{rst}_k[\mu(e_i, e_j,
e_r)\wedge e_s\wedge e_t+e_r\wedge\mu(e_i, e_j, e_s)\wedge
e_t+e_r\wedge e_s \wedge\mu(e_i, e_j, e_t)]$

\vspace{2mm}\noindent $+\sum\limits_{r<s<t}a^{rst}_i[\mu(e_j, e_k,
e_r)\wedge e_s\wedge e_t+e_r\wedge\mu(e_j, e_k, e_s)\wedge
e_t+e_r\wedge e_s \wedge\mu(e_j, e_k, e_t)]$

\vspace{2mm}\noindent $+\sum\limits_{r<s<t}a^{rst}_j[\mu(e_k, e_i,
e_r)\wedge e_s\wedge e_t+e_r\wedge\mu(e_k, e_i, e_r)\wedge
e_t+e_r\wedge e_s \wedge\mu(e_k, e_i, e_t)]$

\vspace{2mm}\noindent$=\sum\limits_{r<s<t}\sum\limits_{l=1}^ma^{rst}_k[c_{ijr}^l
e_l\wedge e_s\wedge e_t+c_{ijs}^l e_r\wedge e_l\wedge e_t+c_{ijt}^l
e_r\wedge e_s \wedge e_l]$

\vspace{2mm}\noindent
$+\sum\limits_{r<s<t}\sum\limits_{l=1}^ma^{rst}_i[c_{jkr}^le_l\wedge
e_s\wedge e_t+c_{jks}^le_r\wedge e_l\wedge e_t+c_{jkt}^le_r\wedge
e_s \wedge e_l]$

\vspace{2mm}\noindent
$+\sum\limits_{r<s<t}\sum\limits_{l=1}^ma^{rst}_j[c_{kir}^le_l\wedge
e_s\wedge e_t+c_{kis}^le_r\wedge e_l\wedge e_t+c_{kit}^le_r\wedge
e_s \wedge e_l].$\hfill(3.10)

\vspace{2mm}\noindent Comparing the identities (3.9) and (3.10), we
obtain

\vspace{2mm}\noindent$a^{rst}_k(c_{ijr}^r+c_{ijs}^s+c_{ijt}^t)+a^{rst}_i(c_{jkr}^r+c_{jks}^s+c_{jkt}^t)+a^{rst}_j(c_{kir}^r+c_{kis}^s+c_{kit}^t)$

\vspace{2mm}\noindent$=\sum\limits_{l=1}^mc_{ijk}^la^{rst}_l,
~~1\leq i<j<k\leq m, ~1\leq r<s<t\leq m;$\hfill(3.11)

\vspace{2mm}\noindent$\sum\limits_{l\neq r,s,t
}[c_{ijr}^la^{rst}_k+c_{kir}^la^{rst}_j+c_{jkr}^la^{rst}_i]=0,
~~1\leq i<j<k\leq m, ~1\leq r<s<t\leq m;$\hfill(3.12)

\vspace{2mm}\noindent$\sum\limits_{l\neq r,s,t
}[c_{ijs}^la^{rst}_k+c_{kis}^la^{rst}_j+c_{jks}^la^{rst}_i]=0,
~~1\leq i<j<k\leq m, ~1\leq r<s<t\leq m;$\hfill(3.13)

\vspace{2mm}\noindent$\sum\limits_{l\neq r,s,t
}[c_{ijt}^la^{rst}_k+c_{kit}^la^{rst}_j+c_{jkt}^la^{rst}_i]=0,
~~1\leq i<j<k\leq m, ~1\leq r<s<t\leq m.$\hfill(3.14)

Conversely, if a $3$-Lie algebra $(L, \mu)$ and a $3$-Lie coalgebra
$(L, \Delta)$ defined  by $(3.8)$, respectively, satisfy identities
(3.11)-(3.14), then $\mu, \Delta$ satisfy identity (3.1).  This
proves the following result.

\vspace{2mm}\noindent {\bf Theorem 3.5 } Let $ L$ be a vector space
with a basis $e_1, \cdots e_m$, $(L, \mu)$ and $(L, \Delta)$ be
$3$-Lie algebra and  $3$-Lie coalgebra  defined by (3.8). Then $(L,
\mu, \Delta)$ is a $3$-Lie bialgebra if and only if $c_{ijk}^l$ and
$a^{rst}_l$, $~ 1\leq i< j<k\leq m,  1\leq l\leq m, $ satisfy
identities (3.11)-(3.14). ~~ $\Box$

\vspace{2mm}\noindent{\bf  Definition 3.6 } Two $3$-Lie bialgebras
$(L_1, \mu_1, \Delta_1)$ and $(L_2, \mu_2, \Delta_2)$ are called
equivalent if there exists a vector space isomorphism $f:
L_1\rightarrow L_2$ such that

\vspace{2mm}\noindent(1) $f: (L_1, \mu_1)\rightarrow(L_2, \mu_2)$ is
a 3-Lie algebra isomorphism, that is,

\vspace{2mm}\noindent $f\mu_1(x,y,z)=\mu_2(f(x),f(y),f(z))$ for
$\forall x,y,z\in L_1;$

\vspace{2mm}\noindent (2) $f: (L_1, \Delta_1)\rightarrow(L_2,
\Delta_2)$ is a 3-Lie coalgebra isomorphism, that is,

\vspace{2mm}\noindent$\Delta_2(f(x))=(f\otimes f\otimes
f)\Delta_1(x)$ for every $ x\in L_1.$ $\Box$

For a given $3$-Lie algebra $L$, in order to find all the $3$-Lie
bialgebra structures on $L$, we should find all the $3$-Lie
coalgebra structures on $L$ that are compatible with the 3-Lie
algebra $L$. Although a permutation of the basis of  $L$ gives the
equivalent $3$-Lie coalgebra structure on $L$, it may leads to a
different 3-Lie bialgebra structure on $L$.

\vspace{2mm}\noindent {\bf Example 3.7 } Let $L$ be a
$4$-dimensional $3$-Lie algebra with a basis $e_1, e_2, e_3, e_4,$
and the multiplication   $\mu: L\otimes L\otimes L \rightarrow L$ as
 $\begin{cases}
\mu(e_2, e_3, e_4)=e_1,\\
\mu(e_1, e_3, e_4)=e_2.\\
\end{cases} $ ~ Defining the three linear mappings $\Delta_1, \Delta_2,
\Delta_3: L\rightarrow L\otimes L\otimes L$ as follows

\vspace{2mm}\noindent$\begin{cases}
\Delta_1 e_1=e_1\wedge e_2\wedge e_4,\\
\Delta_1 e_3=e_3\wedge e_2\wedge e_4,\\
\Delta_1 e_2=\Delta_1 e_4=0;\\
\end{cases}$
~ $\begin{cases}
\Delta_2 e_1=e_1\wedge e_4\wedge e_2,\\
\Delta_2 e_3=e_3\wedge e_4\wedge e_2,\\
\Delta_2 e_2=\Delta_2 e_4=0;\\
\end{cases}$ ~ $\begin{cases}
\Delta_3 e_2=e_2\wedge e_3\wedge e_1,\\
\Delta_3 e_4=e_4\wedge e_3\wedge e_1,\\
\Delta_3 e_1=\Delta_3 e_3=0;\\
\end{cases}$

\vspace{2mm}\noindent we obtain three isomorphic $3$-Lie coalgebras
$(L, \Delta_1)$, $(L, \Delta_2)$, $(L, \Delta_3)$.

By the direct computation, $(L, \mu, \Delta_1)$, $(L, \mu, \Delta_2)
$ and $(L, \mu, \Delta_3)$ are $3$-Lie bialgebras. Let $f: L
\rightarrow L$ be a linear isomorphism defined by:

\vspace{2mm}\noindent $f(e_1)= e_2, ~ f(e_2)= -e_1, ~ f(e_3)= e_4, ~
f(e_4)=e_3.$

\vspace{2mm}\noindent Then $(L, \mu, \Delta_1)$ and $(L, \mu,
\Delta_3)$ are equivalent $3$-Lie bialgebras in the isomorphism $f:
(L, \mu, \Delta_1)\rightarrow(L, \mu, \Delta_3)$. But  $(L, \mu,
\Delta_2) $ is not equivalent to $(L, \mu, \Delta_1)$. ~~ $\Box$

  \vspace{2mm}\noindent{\bf 4. Classification of $3$-dimensional $3$-Lie bialgebras }

In this section we classify the $3$-dimensional $3$-Lie bialgebras
over an algebraically closed  field $F$ of characteristic zero.

\vspace{2mm}\noindent{\bf Lemma 4.1}$^{\small\cite{BSZL}}$ ~ Let
$(L, \mu)$ be an $m$-dimensional $3$-Lie algebra with a basis $e_1,
\cdots, e_m$, $m\leq 4$. Then up to isomorphisms there is one and
only one of the following possibilities:

\vspace{2mm}\noindent(1)  $\dim L \leq 2$, $L$ is abelian, that is,
$\mu=0.$

\vspace{2mm}\noindent(2) $\dim L =3$, $L_a$ is abelian; ~ $L_b$:
$\mu_b(e_1, e_2,e_3)=e_1$.

\vspace{2mm}\noindent(3)  $\dim L=4$,  $L_a$: $L$ is  abelian;
$L_{b_1}:$ $\mu_{b_1}(e_2, e_3,e_4)=e_1$; ~$L_{b_2}:$
$\mu_{b_2}(e_1, e_2,e_3)=e_1$;

\vspace{2mm}\noindent $L_{c_1}\begin{cases}
\mu_{c_1}(e_2, e_3,e_4)=e_1,\\
\mu_{c_1}(e_1, e_3,e_4)=e_2;\\
\end{cases} L_{c_2}\begin{cases}
\mu_{c_2}(e_2, e_3,e_4)=\alpha e_1+e_2,\\
\mu_{c_2}(e_1, e_3,e_4)=e_2, ~\alpha\in F, \alpha\neq 0;\\
\end{cases}$

\vspace{2mm}\noindent  $L_{c_3}\begin{cases}
\mu_{c_3}(e_1,e_3,e_4)=e_1,\\
\mu_{c_3}(e_2, e_3,e_4)=e_2;\\
\end{cases}$
$L_d \begin{cases}
\mu_{d}(e_2, e_3,e_4)=e_1,\\
\mu_d(e_1, e_3,e_4)=e_2,\\
\mu_d(e_1, e_2,e_4)=e_3;\\
\end{cases} L_e\begin{cases}
\mu_e(e_2, e_3,e_4)= e_1,\\
\mu_e(e_1, e_3,e_4)=e_2,\\
\mu_e(e_1, e_2,e_4)=e_3,\\
\mu_e(e_1, e_2,e_3)=e_4.\\
\end{cases}$ ~ $\Box$

\vspace{2mm} Following Theorem 2.3, Theorem 2.4 and Lemma 4.1, we
have the classification of $m$-dimensional $3$-Lie coalgebras with
$m\leq 4$.

\vspace{2mm}\noindent{\bf Lemma 4.2 }  Let $(L, \Delta)$ be an
$m$-dimensional $3$-Lie coalgebra with a basis $e^1,$ $ \cdots, $
$e^m,$ $m\leq 4.$ Then up to isomorphisms there is one and only one
of the following possibilities:

 \vspace{2mm}\noindent(1)  $\dim L\leq 2$,  $(L, \Delta_a)$ is
trivial, that is,  $\Delta_a=0$.

\vspace{2mm}\noindent(2) $\dim L=3$, $L$ is trivial with $\Delta=0$;
$C_b$: $\Delta_b(e^1)=e^1\wedge e^2\wedge e^3.$

\vspace{2mm}\noindent(3) $\dim L=4$, $C_a:$ $(L, \Delta_a)$ is
trivial;

\vspace{4mm}\noindent $C_{b_1}:  \Delta_{b_1}(e^1)=e^2\wedge
e^3\wedge e^4;$ ~ $C_{b_2}:
 \Delta_{b_2}(e^1)=e^1\wedge e^2\wedge e^3; $

\vspace{2mm}\noindent $ C_{c_1}  \begin{cases} \Delta_{c_1}(e^1)=e^2\wedge e^3\wedge e^4,\\
\Delta_{c_1}(
e^2)=e^1\wedge e^3\wedge e^4;\\
\end{cases}$$ C_{c_2}  \begin{cases}
\Delta_{c_2}(e^1)=\alpha e^2\wedge e^3\wedge e^4,  \alpha\in F,  \alpha\neq0,\\
\Delta_{c_2}(e^2)=e^2\wedge e^3\wedge e^4+e^1\wedge e^3\wedge e^4;\\
\end{cases}$

\vspace{2mm}\noindent$ C_{c_3} \begin{cases} \Delta_{c_3}(e^1)=e^1\wedge e^3\wedge e^4,\\
\Delta_{c_3}(e^2)=e^2\wedge e^3\wedge e^4;\\
\end{cases}$$ C_{d} \begin{cases} \Delta_{d}(e^1)=e^2\wedge e^3\wedge e^4,\\  \Delta_{d}(e^2)=e^1\wedge
e^3\wedge e^4,\\ \Delta_{d}(e^3)=e^1\wedge e^2\wedge e^4;\\
\end{cases}$$ C_{e} \begin{cases} \Delta_{e}(e^1)=e^2\wedge e^3\wedge e^4,\\\Delta_e(e^2)=e^1\wedge
e^3\wedge e^4,\\ \Delta_e(e^3)=e^1\wedge e^2\wedge e^4,\\\Delta_e(e^4)=e^1\wedge e^2\wedge e^3.\\
\end{cases}$

\vspace{2mm}\noindent {\bf Proof } The results follow from Theorem
2.3, Theorem 2.5 and Lemma 3.1, directly. ~~ \hfill$\Box$

\vspace{2mm}\noindent{\bf Theorem 4.3 }  Let $(L, \mu, \Delta)$ be a
$3$-dimensional $3$-Lie bialgebra, then $L$ is equivalent to one and
only one of the following posibilities:

 \vspace{2mm}\noindent$(L, 0, 0),$ that is, $\mu=0$ and $\Delta=0$;

\vspace{2mm}\noindent $(L, 0, \Delta_1)$:  $\mu=0, \Delta_1
e_1=e_1\wedge e_2\wedge
 e_3;
$

\vspace{2mm}\noindent $(L, \mu_b, 0):~ \Delta=0;$

 \vspace{2mm}\noindent$(L, \mu_b, \Delta_1)$:
$ ~\Delta_1 e_1=e_1\wedge e_2\wedge e_3;$

\vspace{2mm}\noindent $(L, \mu_b, \Delta_2)$:
$\Delta_2(e_2)=e_2\wedge e_3\wedge e_1;$

\vspace{2mm}\noindent$(L, \mu_b, \Delta_3)$:
$\Delta_3(e_1)=e_1\wedge e_3\wedge e_2,$

\vspace{2mm}\noindent where $\mu_b$ is defined in Lemma 4.1.

\vspace{2mm}\noindent{\bf Proof }  If $(L, \mu)$ is the abelian
$3$-Lie algebra, then by Lemma 4.1 and Lemma 4.2, non-equivalent
3-Lie bialgebras are only $(L, 0, 0)$ and $(L, 0, \Delta_1)$.

If $(L, \mu)$ is a $3$-Lie algebra with $\mu=\mu_b$, then by the
direct computation, the possibilities of $3$-Lie bialgebras $(L,
\mu_b, \Delta_i)$ are as follows: $(L, \mu_b, 0)$;

\vspace{2mm}\noindent $(L, \mu_b, \Delta_1)$:
$\Delta_1(e_1)=e_1\wedge e_2\wedge e_3, ~ \Delta_1(e_2)=\Delta_1
e_3=0;$

\vspace{2mm}\noindent$(L,  \mu_b, \Delta_2)$:
$\Delta_2(e_2)=e_2\wedge e_3\wedge e_1, ~ \Delta_2(e_1)=\Delta_2
e_3=0;$

\vspace{2mm}\noindent $(L, \mu_b, \Delta_3)$:
$\Delta_3(e_1)=e_1\wedge e_3\wedge e_2, ~ \Delta_3(e_2)=\Delta_3
e_3=0;$

\vspace{2mm}\noindent $(L, \mu_b, \Delta_4)$:
$\Delta_4(e_2)=e_2\wedge e_1\wedge e_3, ~ \Delta_4(e_1)=\Delta_4
e_3=0;$

\vspace{2mm}\noindent $(L, \mu_b, \Delta_5)$:
$\Delta_5(e_3)=e_3\wedge e_1\wedge e_2, ~ \Delta_5(e_1)=\Delta_5
e_2=0;$

\vspace{2mm}\noindent $(L, \mu_b, \Delta_6)$:
$\Delta_6(e_3)=e_3\wedge e_2\wedge e_1, ~ \Delta_6(e_1)=\Delta_6
e_2=0.$

\vspace{2mm} Defines linear mappings $f_i: L\rightarrow L$ for $i=1,
2$ as follows:

\vspace{2mm}\noindent$f_1(e_1)=e_1,  f_1(e_2)=e_3, f_1(e_3)=-e_2;$~
$f_2(e_1)=e_1,  f_2(e_2)=-e_2, f_2(e_3)=-e_3.$

\vspace{2mm} Then by the direct computation $(L, \mu_b, \Delta_2)$
is equivalent to $(L, \mu_b, \Delta_5)$, $(L, \mu_b, \Delta_4)$ is
equivalent to $(L, \mu_b, \Delta_6)$ in the mapping $f_1$,
respectively. $(L, \mu_b, \Delta_2)$ is equivalent to $(L, \mu_b,
\Delta_4)$ in the mapping $f_2$. And $(L, \mu_b, \Delta_i)$ is not
equivalent to $(L, \mu_b, \Delta_j)$ for $1\leq i\neq j\leq 3.$
~~$\Box$

\vspace{10mm}\centerline{\textbf{ References }}
\newcounter{fig}
\begin{list}{\upshape[\arabic{fig}]}
{\usecounter{fig}
\setlength{\labelsep}{1mm}\setlength{\labelwidth}{1cm}\setlength{\leftmargin}{1.1cm}}

\bibitem{F} V  Filippov. $n$-Lie algebras,  {\it Sib. Mat. Zh.,}  26 (1985) 126-140.

\bibitem{BL} J. Bagger, N. Lambert, Gauge symmetry
           and supersymmetry of multiple $M2$-branes, {\it Phys. Rev. D,}
           77 (2008) 065008.

\bibitem{HHM} P. Ho, R. Hou and Y. Matsuo, Lie $3$-algebra and multiple
$M_2$-branes, {\it JHEP, } 0806 (2008) 020.

\bibitem{HCK} P. Ho, M. Chebotar and W. Ke, On skew-symmetric maps on
Lie algebras, {\it Proc. Royal Soc. Edinburgh A,} 113 (2003)
1273-1281.

\bibitem{G} A. Gustavsson, Algebraic structures on parallel
M2-branes, {\it Nucl.Phys,} B811 (2009) 66-76.

\bibitem{P} G. Papadopoulos, M2-branes, $3$-Lie algebras and
Plucker relations, {\it JHEP,} 0805 (2008) 054.

\bibitem{K} S. Kasymov, On a theory of $n$-Lie algebras, {\it Algebra i Logika,} 26 (1987)
277-297.

\bibitem{L} W. Ling, On the structure of $n-$Lie
                  algebras, Dissertation, {\it University-GHS-Siegen, Siegn,} 1993.

\bibitem{Pa} A. P. Pojidaev, Enveloping algebras of Filippov algebras, {\it Comm. Algebra,} 31(2) (2003): 883-900.

\bibitem{Pb}  A. P. Pozhidaev,  Solvability of finite -
             dimensional n-ary commutative Leibniz algebras of characteristic
             0, {\it Comm. Algebra,} 31(1) (2003): 197-215.

\bibitem{YLZ} Y. Jin, W. Liu, and Z. Zhang, Real simple n-lie
algebras admitting metric structures, {\it J. Phys. A: Math. Theor,}
42 (2009) 9pp

\bibitem{BSZ}  R. Bai , C. Shen  and Y. Zhang, 3-Lie algebras with an ideal N,
{\it Linear Alg. Appl,} 431 (2009) 673-700.

\bibitem{BBW} R Bai, C Bai and J Wang.  Realizations of $3$-Lie
algebras, {\it J. Math. Phys,} 2010, 51: 063505.

\bibitem{BSZL} R. Bai, G. Song and Y. Zhang, On classification of n-Lie
algebras, {\it Front. Math. China},   6 (4) (2011) 581-606.

\bibitem{M}  W. Michaelis, Lie coalgebra, {\it Adv. Math,}  38 (1980):1-54.

\bibitem{BD}  A. A. Bolavin, V.G. Drinfeld, Solutions of the classical
Yang- Baxter equation for simple Lie algebras, {\it Func. Anal.
Appl}, 16 (1982):159-180.

\bibitem{VD} V. DeSmedt,  Existence of a Lie bialgebra structure on every
Lie algebra, {\it Lett. Math. Phys}, 31(1994): 225-231.

\end{list}
\end{document}